\documentclass{iaus}
\usepackage{graphicx}

\title[NBursts: Kinematics and SFH from integrated spectra] 
{NBursts: Simultaneous Extraction of Internal
Kinematics and Parametrized SFH from Integrated Light Spectra}

\author[Chilingarian et al.]   
{Igor Chilingarian$^{1,2,3}$%
  \thanks{e-mail: chil@sai.msu.su},
 Philippe Prugniel$^{2,4}$,
 Olga Sil'chenko$^{1}$ \and
 Mina Koleva$^{2,5}$
}

\affiliation{$^1$Sternberg Astronomical Institute, Moscow State University,
13 Universitetsky prospect, Moscow, 119992, Russia;\\[\affilskip]
$^2$Centre de Recherche Astronomique de Lyon, Observatoire de Lyon, 9 Av. Charles 
Andr\'e, Saint-Genis Laval, F-69561, France; CNRS, UMR 5574;\\[\affilskip]
$^3$Observatoire de Paris, LERMA, 61 Ave. de l'Observatoire, Paris, 75014, France;\\[\affilskip]
$^4$Observatoire de Paris-Meudon, GEPI, 9 pl. Jules Janssen, Meudon, 92195, France;\\[\affilskip]
$^5$Department of Astronomy, St. Kliment Ohridski University of Sofia, 5 James Bourchier Blvd., BG-1164 Sofia, Bulgaria
}

\pubyear{2007}
\volume{241}  
\pagerange{1-2}
\date{30 Jan 2007}
\jname{Proceedings Stellar Populations as Building Blocks of Galaxies}
\editors{A. Vazdekis \& R. Peletier, eds.}
\begin{document}

\maketitle

\begin{abstract}
We present a novel approach for simultaneous extraction of stellar
population parameters and internal kinematics from the spectra integrated
along a line of sight. We fit a template spectrum into an observed one
in a pixel space using a non-linear $\chi^2$ minimization in the
multidimensional parameter space, including characteristics of the
line-of-sight velocity distribution (LOSVD) and parametrized star formation
history (SFH). Our technique has been applied to IFU and multi-object
spectroscopy of low-luminosity early type galaxies.

\end{abstract}

\firstsection 
\section{Introduction}
In a ``classical'' way, extraction of stellar population parameters and
internal kinematics from absorption-line spectra of galaxies, integrated
along a line of sight go independently. Internal kinematics is determined by
deconvolving the observed spectrum with a single spectrum of a star
(``template spectrum''), observed with the same setup (see e.g. Sargent et
al. 1977), or with a linear combination of template spectra (optimal
template fitting) (van der Marel \& Franx, 1993). Stellar population
parameters are normally estimated using absorption line strength indices:
low-resolution ($\sim$10\AA) Lick indices (Worthey et al. 1994), or
intermediate-resolution indicators (e.g. Vazdekis \& Arimoto, 1999). 

Caveats of the classical approaches are: (1) the template mismatch, putting
limitations on the precision of kinematics, (2) non-optimal usage of
information, contained in the spectrum, when measuring absorption line
strength indices. One of the modern approaches, STECKMAP (Ocvirk et al.
2006) performs analysis of the complete spectral energy distribution to
extract simultaneously SFH and LOSVD in a non-parametric way.

\section{The NBursts method and its applications}
We propose a new approach for determination of parametrized LOSVD and SFH.
Our method is the generalization of the penalized pixel fitting algorithm
(Cappellari \& Emsellem, 2004). An optimal template is represented by the
linear combination of SSP's with free ages and metallicities, determined in
the same minimization loop, using interpolation of the pre-computed grid of
PEGASE.HR SSP's (Le Borgne et al. 2004).

The $\chi^2$ value (without penalization) is computed as follows:
\begin{eqnarray}
        \chi^2 = \sum_{N_{\lambda}}\frac{(F_{i}-P_{1p}(T_{i}(SFH) \otimes
        \mathcal{L}(v,\sigma,h_3,h_4) + P_{2q}) )^2}{\Delta F_{i}^2}, 
        \nonumber\\
        \mbox{where} \quad T_{i}(SFH) = \sum_{N_{bursts}}k_{i} T_{i}(t_n, Z_n)
\label{chi2eq}
\end{eqnarray}

where $\mathcal{L}$ is LOSVD; $F_{i}$ and $\Delta F_{i}$ are observed flux
and its uncertainty; $T_{i}(SFH)$ is the flux from an synthetic spectrum,
represented by a linear combination of $N_{bursts}$ SSP's and convolved
according to the line-spread function of the spectrograph; $P_{1p}$ and
$P_{2q}$ are multiplicative and additive Legendre polynomials of orders $p$
and $q$ for correcting the continuum; $t$ is age, $Z$ is metallicity, $v$,
$\sigma$, $h_3$, and $h_4$ are radial velocity, velocity dispersion and
Gauss-Hermite coefficients respectively (Van der Marel \& Franx, 1993). 
Precision of the parameter estimates with our technique are discussed in
details in Koleva et al. 2006.

The simplest case, $N_{burst}=1$, allows to compare of with the stellar
population parameters derived using Lick indices. Due to optimized usage
of information contained in the spectra, our technique provides 2 to 5 time
better precision compared to Lick indices for the same S/N ratio for a
typical spectral range and resolution of modern intermediate-resolution
instruments: 4200\AA$<\lambda<$5600\AA; $R=1800$.

We used NBursts to analyse data on early type dwarf galaxies: (1) 3D
spectroscopy of dE galaxies in the Virgo cluster and nearby groups (MPFS IFU
spectrograph at the Russian 6-m telescope), (2) high-resolution multi-object
spectroscopy of a sample of early-type galaxies in the Abell 496 cluster
(FLAMES-Giraffe at ESO VLT).

Details on the results and interpretation are available in Chilingarian
(2006, PhD thesis). Kinematically (Chilingarian et al. 2007) and
evolutionary (Chilingarian et al. 2006) decoupled structures are revealed in
most of the galaxies from our 3D spectroscopic sample. Disky features and
intermediate-age stellar population strengthen the connection between early
and late-type dwarf galaxies

\begin{acknowledgments}
We are very grateful to the financial support, provided by the IAU to IC and
MK. PhD thesis of IC is supported by the INTAS YS grant 04-83-3618, Studies
of dE galaxies at SAI MSU are supported by the bilateral Russian-Flemish
grant RFBR 05-02-19805MF\_a.
\end{acknowledgments}

\end{document}